\documentclass[aip,jcp,reprint]{revtex4-1}
\pdfoutput=1
\usepackage{graphicx}
\usepackage{amssymb}
\usepackage{amsmath}
\usepackage{bm}
\usepackage{microtype}
\usepackage{color}

\newcommand{\bra}[1]{\langle #1 |}
\newcommand{\ket}[1]{| #1 \rangle}

\newcommand{\ep}{\epsilon}

\newcommand{\bmr}{\mathbf{r}}

\newcommand{\eqr}[1]{Eq.~\eqref{eq:#1}}

\newcommand{\eql}[1]{\label{eq:#1}}
\newcommand{\figr}[1]{Fig.~\ref{fig:#1}}
\newcommand{\figl}[1]{\label{fig:#1}}

\newcommand{\tcb}[1]{\textcolor{blue}{#1}}
\newcommand{\tcr}[1]{\textcolor{red}{#1}}
\newcommand{\bse}{\begin{subequations}}
\newcommand{\ese}{\end{subequations}}

\begin{document}
\bibliographystyle{tim}

\title{Theoretical description of Carbene-Metal-Amides}
\author{Timothy J. H. Hele}
\email{tjhh2@cam.ac.uk}
\author{Dan Credgington}
\affiliation{Cavendish Laboratory, JJ Thomson Avenue, Cambridge University, CB3 0HE, UK.}

\begin{abstract}
Carbene-Metal-Amide light-emitting diodes have recently shown internal quantum efficiencies approaching 100\%, and there has been substantial debate concerning the cause of their exceptionally high efficiency. 
Here we present a theoretical description of CMAs, showing how a simple three-atom model can predict the form of the HOMO and LUMO, determine the polarization of transitions and the feasibility of spin-orbit coupling, as well as the qualitative dependence of excited state energies and oscillator strength on the twist angle. These results clarify many of the claims concerning CMAs and pave the way for the design of more efficient devices.
\end{abstract}

\maketitle

\section{Introduction}
Carbene-metal-amides (CMAs) have recently emerged as candidates for next-generation organic light-emitting diodes (OLEDs) with internal quantum efficiencies far exceeding the spin statistics threshold of 25\%\cite{di17a}. 
In a recent article\cite{di17a}, Di \emph{et al.} suggested that the cause of this surprisingly high efficiency was an inversion in the energies of the lowest excited singlet ($S_1$) and lowest excited triplet ($T_1$), as evidenced by an unusual ordering of the phosphorescence and (delayed) fluorescence energies.  DFT and TDDFT calculations suggested rotation around the intramolecular CMA  bond lowered the energy of $S_1$ to the extent that it crossed the energy of $T_1$, facilitating rapid reverse intersystem crossing from a dark $T_1$ to emissive $S_1$ and thereby allowing triplets to emit.

Shortly afterwards, F\"oller and Marian\cite{fol17a} suggested that the rotationally-accessed spin-state inversion (RASI) postulated by Di \emph{et al.}\ was an artifact of their calculations, and that the experimentally observed photophysics could instead be explained by considering solvent reorganization effects. Like Di \emph{et al.}, they found that triplet/singlet interconversion was rapid under thermal conditions, but did not find inversion of the $S_1$ and $T_1$ energies.\cite{fol17a} These conclusions were based on multiconfigurational calculations using a parameterized, semiempirical method, which unlike TDDFT can include double and higher excitations.\cite{fol17a}

The purpose of this letter is not to definitively determine the mechanism of action of CMAs but to provide a consistent theoretical framework in which to interpret the ostensibly conflicting results. Given that extensive experimental and computational results are available, here we consider the model theoretically and show how by combining molecular orbital theory and group theory, a three-atom model can explain many of the computed and experimentally-observed properties of these systems. We also use electronic structure theory to discuss the necessary conditions for RASI to occur. It is hoped that this will provide a unified framework for the rational design of future CMAs and determining their mechanism of action.

\section{Electronic structure theory}
We begin by considering the energy ordering of the first excited singlet and triplet states. Strictly speaking, the $S_1$ and $T_1$ electronic wavefunctions are given by configuration interaction expansions\cite{sza89a} which are linear combinations of excitations that diagonalize the electronic (Born-Oppenheimer) Hamiltonian $\hat H_{\rm el}$. Both Di \emph{et al.}\ and F\"oller and Marian found the $S_1$ and $T_1$ states to be dominated by excitation from the highest occupied molecular orbital (HOMO) to the lowest unoccupied molecular orbital (LUMO). We formally denote these HOMO$\rightarrow$LUMO excitations $^1\Psi_H^L$ and $^3\Psi_H^L$ respectively where the $^1$ and $^3$ superscripts correspond to singlet and triplet spin-adapted linear configurations\cite{sza89a}. From standard electronic structure theory\cite{sza89a} we obtain the exact results
\begin{align}
 E(^1\Psi_H^L) = \bra{^1\Psi_H^L} \hat H_{\rm el} \ket{^1\Psi_H^L} = & \ep_H - \ep_L - J_{HL} + 2K_{HL} \\
 E(^3\Psi_H^L) = \bra{^3\Psi_H^L} \hat H_{\rm el} \ket{^3\Psi_H^L} = & \ep_H - \ep_L - J_{HL}
\end{align}
where $\ep_H$ and $\ep_L$ are the energies of the HOMO and LUMO respectively (diagonal elements of the Fock matrix), $J_{HL}$ is the Coulomb attraction of an electron in the LUMO and a hole in the HOMO and $K_{HL}$ is the exchange integral\cite{sza89a},
\begin{align}
 K_{HL} = \int d\bmr_1 \int d\bmr_2 \ \phi_H^*(\bmr_1) \phi_L(\bmr_1) \frac{1}{|\bmr_1-\bmr_2|} \phi_L^*(\bmr_2) \phi_H(\bmr_2) \eql{ex}. 
\end{align}
Sometimes `Exchange energy' is used to mean $E(S_1) - E(T_1)$\cite{di17a}; we refrain from this terminology here to avoid confusion with the exchange integral defined above. These results hold for the exact Born-Oppenheimer electronic Hamiltonian $\hat H_{\rm el}$ but not necessarily at more approximate levels of theory such as TDDFT. We immediately see that the energy separation between these excitations is
\begin{align}
 E(^1\Psi_H^L) - E(^3\Psi_H^L) = 2K_{HL}.
\end{align}
However, exchange integrals are always positive\cite{roo51a} such that $K_{HL} \geq 0 $ and 
\begin{align}
 E(^1\Psi_H^L) - E(^3\Psi_H^L) \geq 0.
\end{align}
We consequently find that, if $\ket{S_1} = \ket{^1\Psi_H^L}$ and $\ket{T_1} = \ket{^3\Psi_H^L}$, as has previously been approximated, then spin-state inversion is impossible, whether assisted by rotation or not. This means that for spin-state inversion to occur $^1\Psi_H^L$ must mix substantially with other, higher-lying excitations (if only $^3\Psi_H^L$ mixed then by the variational principle this would lower its energy and increase the singlet-triplet energy gap). As F\"oller and Marian suggest, this could occur by mixing with double excitations that can only be of singlet character\cite{fol17a}. However, it could also be from direct mixing with single and triple excitations (or double excitations that can be either triplet or singlet) that mix more strongly with $^1\Psi_H^L$ than $^3\Psi_H^L$, since the mixing elements between similar singlet excitations are not necessarily equal to mixing elements between corresponding triplet excitations\cite{par56a}. 


\section{Model three-atom system}
We now construct the simplest model of the CMA chromophore which can describe the relevant photophysics and also be adaptable to a variety of CMA compounds. Considering CMA1 synthesized by Di \emph{et al.}\cite{di17a}, shown in \figr{molstruc}(a), the carbazole $\pi$ system is likely to interact with the metal and carbene, but this is only significantly mediated through its nitrogen atom. The adamantane group is unlikely to contribute to the chromophore's electronic structure since its $\sigma$ bonding orbitals will be too low in energy and its $\sigma^*$ orbitals too high. While the benzene ring moiety probably has orbitals of similar energy to the carbazole unit, its $\pi$ system is orthogonal to that of the nearby N and carbene and therefore not able to conjugate with the carbene, metal or carbazole. The nitrogen adjacent to the carbene is likely to stabilize the carbene and contribute to the electronic structure, but only via the carbene carbon.

We therefore consider a three-atom system, the carbene carbon, the energy of whose vacant $2p$ orbital can implicitly include stabilization from the adjacent nitrogen, the metal atom, and the N of the carbazole, whose occupied $2p$ orbital can be a proxy for the carbazole HOMO. This model is sketched in \figr{molstruc} in the planar (b) and twisted (c) conformations.


\begin{figure}[tb]
 \includegraphics[width=1\columnwidth]{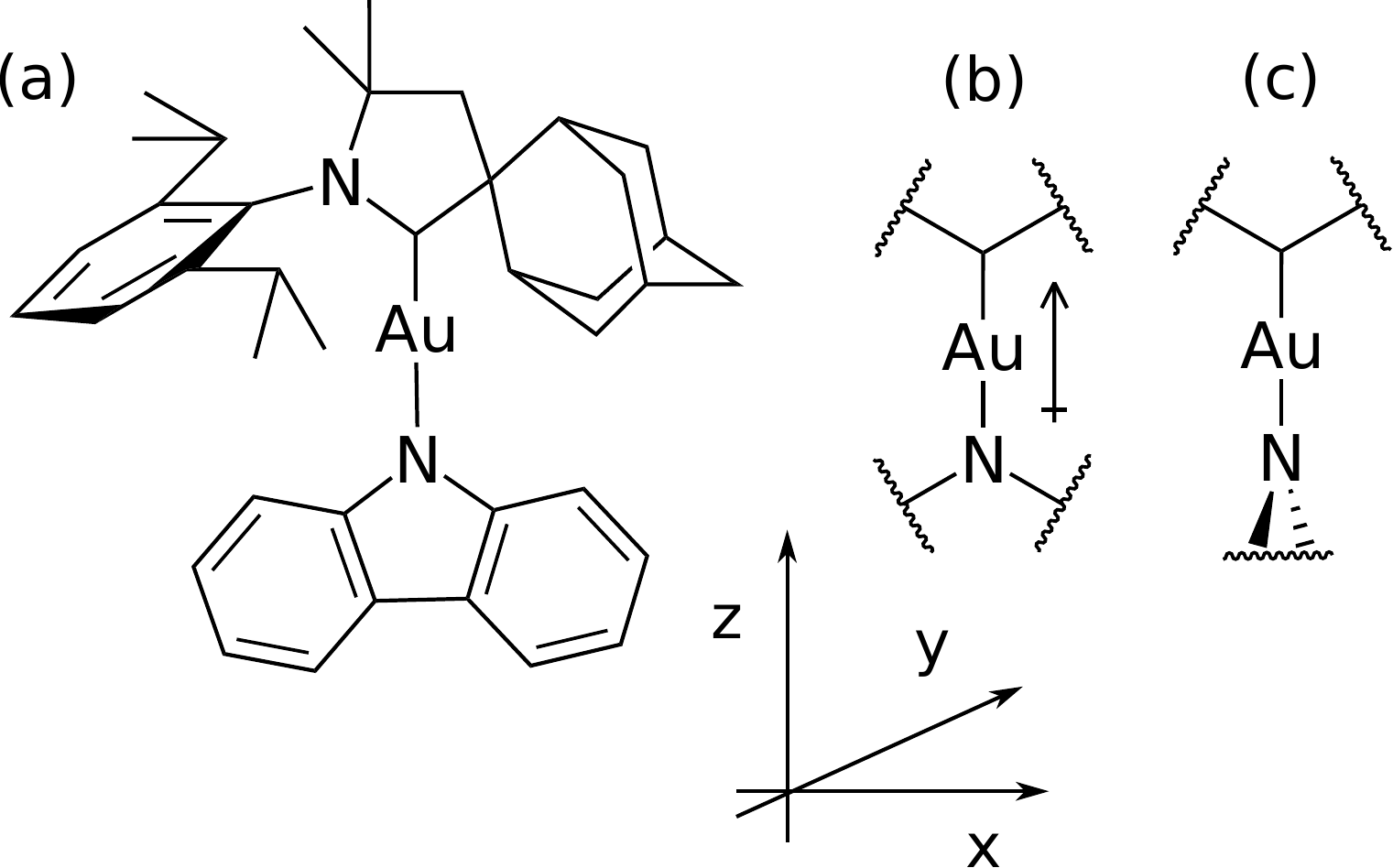}
 \caption{(a) CMA1 structure\cite{di17a} with the simplified three-atom model in the planar (b) and twisted (c) configurations along with the co-ordinate axes used in this article. The permanent ground-state dipole deduced from symmetry and orbital arguments is shown in (b), pointing from negative to positive.}
 \figl{molstruc}
\end{figure}

This model has $C_{2v}$ symmetry\cite{atk11a}, both in the planar and twisted (at 90$^\circ$) geometries, with $C_2$ symmetry at intermediate twist angles. For a `true' chromophore such as CMA1 (which has $C_s$ symmetry), descent in symmetry\cite{atk11a} could be used, though we do not find this necessary in order to explain the observed photophysics of CMA1 except for the nonvanishing $S_1$/$T_1$ spin-orbit coupling (see below). This symmetry assignment allows us to uniquely define the $z$ axis as that of a $C_2$ rotation. We then define the $x$ and $y$ axes such that the planar chromophore lies in the $xz$ plane, and that the carbazole and not the carbene is rotated in the twisted CMA (see axes in \figr{molstruc}). 

We now construct molecular orbitals (MOs) for the CMA by the textbook method of forming a basis, assigning irreducible representations (irreps) to the basis functions, and then considering mixing of basis functions of the same irrep\cite{atk11a}. For consistency with CMA1,\cite{di17a} we consider constructing the chromophore from a neutral (singlet) carbene, Au$^+$ and the carbazole anion (Cz$^{-}$), but this analysis immediately extends to any $d^{10}$ transition metal species such as Cu$^{+}$ used in CMA2,\cite{di17a} and to other amide units such as those in CMA3 and CMA4.\cite{di17a} We stress that the exact form of the MOs and their energies will be a function of the particular CMA, level of calculation, and solvent/solid-state environment, and the purpose here is to develop qualitative MOs to describe the observed photophysics.

Our minimal atomic orbital basis then comprises:
\begin{itemize}
 \item From carbon: carbene lone pair in an $sp^2$-hybridized orbital and an empty $2p$ orbital.
 \item Metal atom: filled $5d$ orbitals, empty $6s$ and empty $6p$.
 \item Nitrogen: anion lone pair in $sp^2$-hybridized orbital, filled $2p$ orbital as a proxy for the carbazole/amide HOMO.
\end{itemize}

We then assign each of these basis functions to irreps of the $C_{2v}$ point group\cite{atk11a}, shown in Table~\ref{tab:irreps}, finding that the only difference between the planar and twisted geometries is the irrep of the nitrogen $2p$ orbital.

\begin{table}[h]
 \begin{tabular}{c|c|c|c|c}
   & C & Au$^+$ & N$^{-}$ (planar) & N$^-$ (twisted) \\ \hline 
   $A_1$ & l.p. & $6s,5d_{z^2},5d_{x^2-y^2},6p_z $& l.p. & l.p. \\
   $A_2$ & & $5d_{xy}$ &  & \\  
   $B_1$ & & $5d_{xz},6p_x$ & & $2p$ \\
   $B_2$ & $2p$ & $5d_{yz},6p_y$ & $2p$& 
 \end{tabular}
\caption{Irreps of the orbitals of the model chromophore in the $C_{2v}$ point group. l.p.\ = $sp^2$-hybridized lone pair.}
\label{tab:irreps}
\end{table}

To determine the MOs, we start with orbitals belonging to the totally symmetric irrep $A_1$, and which are the same for both the planar and twisted geometries. From chemical intuition, the filled (degenerate) $5d_{z^2}$, and $5d_{x^2-y^2}$ will be lower in energy than the empty $6s$, which in turn is lower in energy than the empty $6p_z$. Orbital overlap arguments show that $5d_{x^2-y^2}$ cannot mix appreciably with any other orbitals and is essentially nonbonding. The filled C and N lone pairs will be of similar energy, with N possibly lower than C due to its greater electronegativity. We can consider the filled N and C $sp^2$ orbitals to form in-phase and out-of-phase combinations (corresponding to sign under approximate inversion through the metal atom). The in-phase combination can mix with the $5d_{z^2}$ and $6s$, leading to a 
strongly bonding orbital, a filled nonbonding orbital probably dominated by Au $6s$ and $5d_{z^2}$, and a strongly antibonding orbital. The out-of-phase combination of lone pairs can mix with the Au $6p$ forming bonding and antibonding orbitals. In chemistry terms, this leads to a $\sigma$ bonding system between C--Au--N with two bonding orbitals, one nonbonding orbital and two antibonding orbitals, which holds the CMA together and allows it to twist without disintegrating. The corresponding MO diagram is sketched in \figr{moa1}.

\begin{figure}[tb]
 \includegraphics[width=.65\columnwidth]{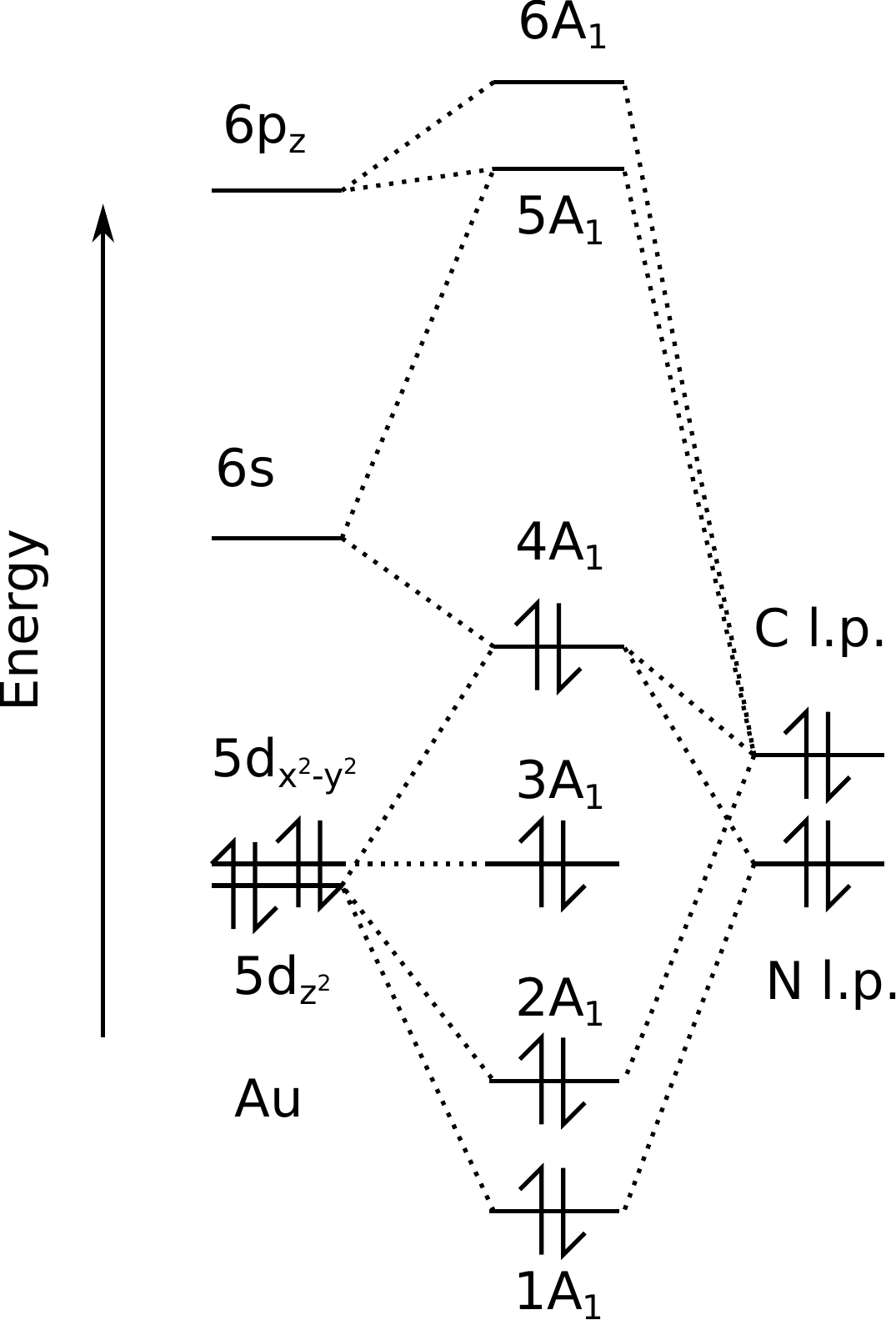}
 \caption{Schematic MO diagram for the $A_1$ irrep of the three-atom model, illustrating the $\sigma$ bonding system in CMAs. For clarity not all interactions between atomic/molecular orbitals are shown. The key finding is that the highest $A_1$ MO is a nonbonding orbital, probably dominated by Au orbitals. l.p.\ = $sp^2$-hybridized lone pair.}
 \figl{moa1}
\end{figure}

The $A_2$ irrep only contains the occupied $5d_{xy}$ orbital which is therefore nonbonding in both the planar and twisted conformations. Even in intermediate twisted conformations it cannot constructively overlap with the N $2p$ since it is of opposite parity under the $C_2$ operation. 

We now form the MOs in the $B_1$ and $B_2$ irreps (\figr{mob1b2}). For a planar conformation the $B_1$ orbitals $5d_{xz}$ (filled) and $6p_x$ (empty) will  be nonbonding [blue orbitals in \figr{mob1b2}(a)]. For $B_2$, The filled $5d_{yz}$ will be of lower energy than the vacant $6p_y$ and the filled N $2p$ orbital of lower energy than the vacant carbene $2p$. We therefore consider the $5d_{yz}$ and N $2p$ to form in-phase and out-of-phase combinations $1B_2$ and $2B_2$ (both of which are occupied) and the $6p_y$ and C $2p$ to form in-phase and out-of-phase combinations $3B_2$ and $4B_2$ (both of which are unoccupied), shown in black in \figr{mob1b2}(a). Qualitatively, the filled N lone pair has pushed up the Au $5d_{yz}$ orbital energy and the empty carbene $p$ orbital has stabilized the $6p_y$. In addition, the $2B_2$ and $3B_2$ orbitals [see \figr{mob1b2}(a)] may mix slightly with each other, with the $2B_2$ acquiring a small bonding interaction with C $2p$ (and falling in energy) and the $3B_2$ acquiring 
an antibonding interaction with the N $2p$ (and rising in energy).  
\begin{figure}[tb]
 \includegraphics[width=\columnwidth]{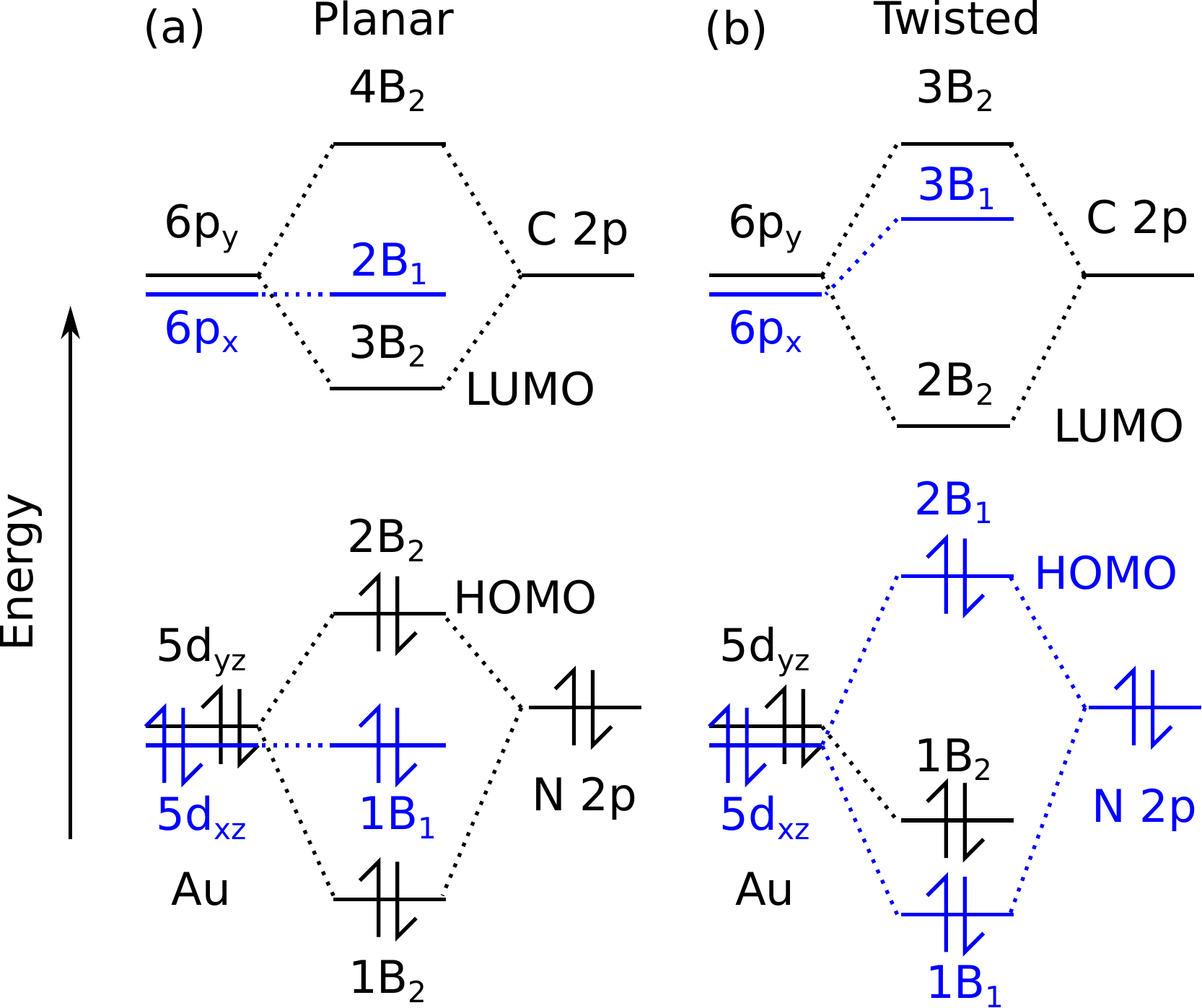}
 \caption{Schematic MO diagram for the $B_1$ (shown in blue) and $B_2$ (shown in black) irreps in the planar (a) and twisted (b) conformation of the CMA, with electronic occupancy in the ground state ($S_0$). The HOMO changes from being $B_2$ to $B_1$ upon rotation and rises slightly in energy; the LUMO remains $B_2$ but falls slightly.}
 \figl{mob1b2}
\end{figure}

For the $B_1$ irrep in twisted geometry, there will be a filled bonding orbital of an in-phase $5d_{xz}$ and some N $2p$, a filled nonbonding orbital which is probably an out-of-phase combination of $5d_{xz}$ and some N $2p$, and an antibonding orbital dominated by  $6p_x$ [blue orbitals in \figr{mob1b2}(b)]. For the $B_2$ irrep, there will be a filled bonding orbital localized on $5d_{yz}$, a non-bonding orbital of the $6p_y$ and C $2p$ in-phase and an antibonding orbital of the $6p_y$ and C $2p$ out-of-phase [black orbitals in \figr{mob1b2}(b)]. 

From these qualitative MO diagrams we see that the in the planar geometry the HOMO is likely to be $2B_2$, and the LUMO $3B_2$. We draw the qualitative form of these orbitals in the $yz$ plane in \figr{hl}(a) and find that they match the HOMO and LUMO of CMA1 as calculated by Di \emph{et al.}\ (Fig.~3B of their article, drawn in approximately the $xz$ plane). The HOMO we predict is somewhat similar to that of F\"oller and Marian (who find amplitude on carbazole but not Au) and to their LUMO (who find amplitude on Au $5d_{yz}$ and C $2p$ but of opposite relative sign). Although neither set of calculations assigns irreps to their states, the qualitative form of their HOMOs and LUMOs are all $B_2$ in the planar geometry. That such a simple three-atom model can qualitatively reproduce the results of some of the electronic structure calculations gives us confidence in using this model to determine the observed photophysics. 

Using similar arguments, in the twisted geometry the HOMO becomes $B_1$ and the LUMO is still $B_2$, which are drawn in \figr{hl}(b). As drawn in \figr{mob1b2}, in the twisted configuration the HOMO/LUMO gap is likely to be smaller as they are no longer of the same irrep and cannot mix with and repel each other. Qualitatively, this corresponds to the HOMO rising in energy and the LUMO falling, explaining why in the ground state ($S_0$) the molecule adopts a planar conformation but in the excited state (where there is one electron in the HOMO and in the LUMO) the potential energy surface is much flatter.\cite{di17a,fol17a}

In both the planar and twisted conformations, the $4A_1$ orbital [\figr{moa1}] is likely to be of comparable energy to the HOMO and participate in the photophysics.

\begin{figure}[tb]
 \includegraphics[width=\columnwidth]{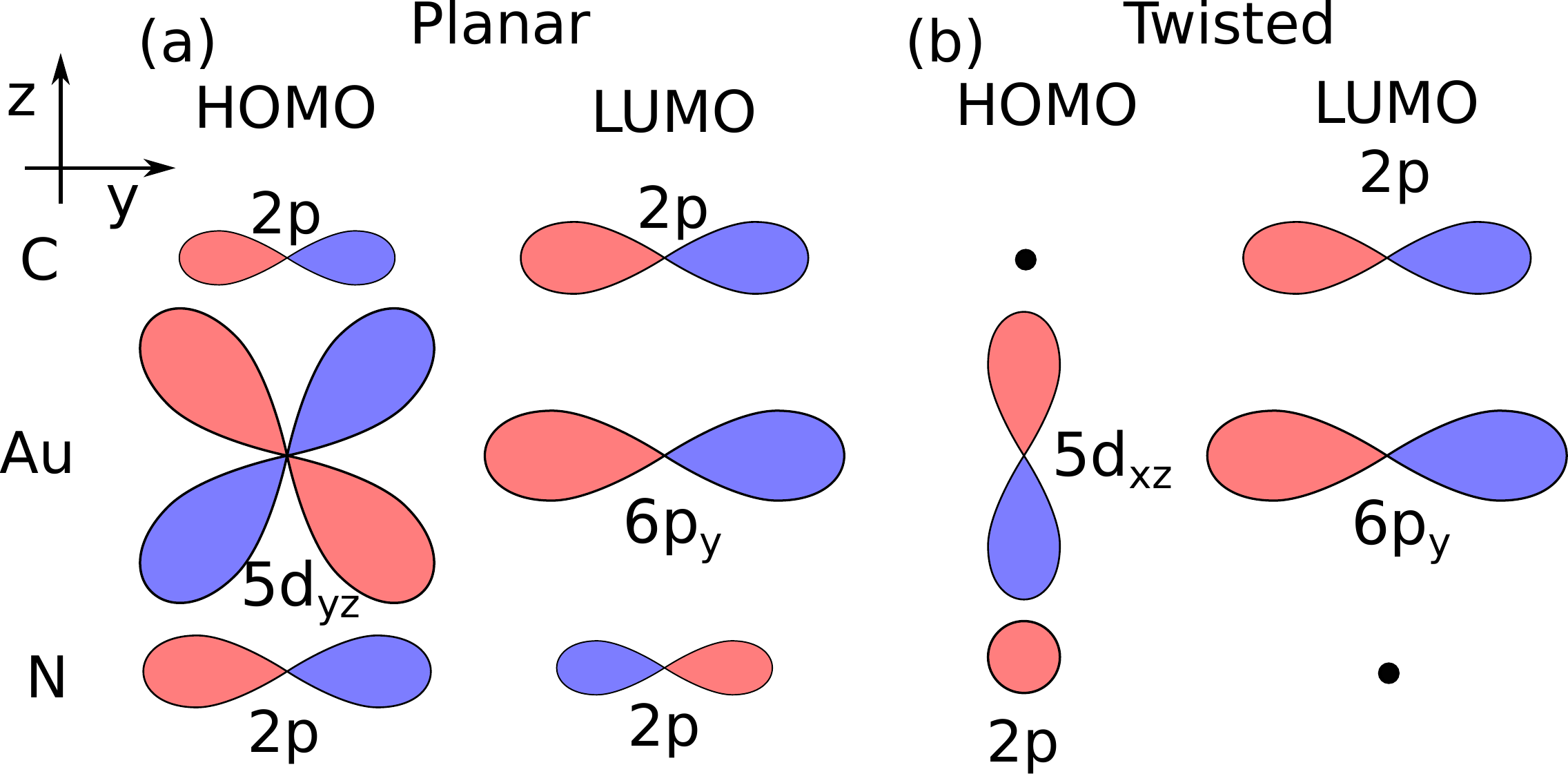}
 \caption{Sketches of the CMA HOMO and LUMO in the planar (left) and twisted (right) conformation, drawn in the $yz$ plane (side on) to show the $\pi$ bonding. A black dot signifies a node. In the planar conformation, the C $2p$ makes a smaller contribution to the HOMO than the N $2p$ and a larger contribution to the LUMO.}
 \figl{hl}
\end{figure}

\section{Photophysics}
For the planar conformation, the HOMO$\to$LUMO excitation transforms as the direct product\cite{atk11a} $B_2 \times B_2 = A_1$, which is dipole-allowed, and polarized in the $z$ direction. Another candidate for a low-energy excitation is $4A_1\to 2B_2$, which transforms as $B_2$, and is again dipole allowed, but $y$ polarized. This appears to be the excitation F\"oller and Marian refer to as $d_{\sigma}\to$LUMO and variously assign to $S_2$ or $S_3$. For consistency, here we define it as $S_2$.\cite{fol17a}

In the twisted geometry, the HOMO$\to$LUMO excitation transforms as $A_2$, which is dipole forbidden, explaining the computational observation by both Di \emph{et al.}\ and F\"oller and Marian that there is vanishing dipole moment in this geometry\cite{di17a,fol17a}. However, the excitation $S_2$ excitation is still allowed, transforming as $B_2$. We summarize these results in Table~\ref{tab:ex}.

\begin{table}[h]
 \begin{tabular}{c|c|c}
  Orbital/excitation & Planar & Twisted \\ \hline
  HOMO--1 & $4A_1$ & $4A_1$ \\
  HOMO & $2B_2$ & $2B_1$ \\
  LUMO & $3B_2$ & $2B_2$ \\
  $S_1$, $T_1$ & $\tcb{A_1}$ & $\tcr{A_2}$ \\
  $S_2$ & $\tcb{B_2}$ & $\tcb{B_2}$
 \end{tabular}
\caption{Irreps of orbitals and excitations mentioned in the text. For singlet excitations, irreps corresponding to dipole-allowed transitions are shown in blue, forbidden in red, and triplet excitations are always forbidden by spin-symmetry. $S_1$ becomes dipole-forbidden upon rotation whereas $S_2$ does not.}
\label{tab:ex}
\end{table}

Nevertheless, satisfying point group symmetry is a necessary but not sufficient criterion for an integral to be nonzero and we therefore explore the origin of the oscillator strength of the $S_0\to S_1$ transition. In the spirit of intermediate neglect of differential overlap\cite{pop67a} we look for atom(s) (if any) upon which there is significant amplitude of both HOMO and LUMO, finding a suitable candidate to be Au, whose $5d_{yz}$ contributes to the HOMO and $6p_y$ contributes to the LUMO. Their overlap consequently leads to the nonzero oscillator strength, such that the metal atom is the \emph{de facto} chromophore of these systems, and the HOMO$\to$LUMO excitation can be likened to a $5d_{yz}\to 6p_y$ Au transition. However, $d \to p$ excitations are usually too high energy to be seen in the visible, but here is red-shifted as the N $2p$ destabilizes the HOMO and the carbene $2p$ stabilizes the LUMO (see above). The oscillator strength of the observed $S_0\to S_1$ transition will, however, still be 
small since most of the HOMO and LUMO are localized outside the metal atom---explaining the experimentally-observed low extinction coefficient of CMAs.\cite{di17a}

\section{Spin-orbit coupling}
The spin-orbit coupling (SOC) operator $\hat H_{\rm so}$ transforms as molecular rotations\cite{atk11a}, which in the $C_{2v}$ point group transform as all irreps except $A_1$. This means that, in order for SOC to mix two states they must be of different symmetry (as qualitatively embodied in El Sayed's rule). However, the electronic wavefunctions of $S_1$ and $T_1$ have the same irrep ($A_1$) so $\bra{S_1} \hat H_{\rm so} \ket{T_1}=0$, explaining the numerically observed small SOC between $S_1$ and $T_1$ seen by F\"oller and Marian\cite{fol17a}. They probably find nonzero $S_1/T_1$ SOC as the true symmetry of CMA1 ($C_s$) is lower than that of the model ($C_{2v}$) (see above). 

Conversely, if we consider SOC between the $S_2$ transition and $T_1$, we find at the planar geometry $B_2\times A_1 = B_2$, the irrep of $R_x$, and at the twisted geometry $B_2\times A_2 = B_1$, the irrep of $R_y$. Consequently, $\bra{S_2} \hat H_{\rm so} \ket{T_1}$ is likely to be nonzero in a range a geometries, consistent with the strong $S_2/T_1$ SOC observed by F\"oller and Marian.\cite{fol17a}

\section{Exchange integral}
While this theoretical analysis cannot definitively determine whether or not spin-state inversion occurs, we can consider how the $S_1/T_1$ energy gap changes upon twisting by examining the exchange integral in \eqr{ex}. As noted by Roothaan\cite{roo51a}, for real orbitals such as those considered here the exchange integral $K_{HL}$ corresponds to the self-interaction energy of the charge distribution $\rho(\bmr) = \phi_H(\bmr)\phi_L(\bmr)$. The HOMO and LUMO only significantly overlap on the metal atom (see above) and  in the planar geometry $\rho_{\rm pl}(\bmr) \simeq \phi_{5d_{yz}}(\bmr)\phi_{6p_y}(\bmr)$, which will have a node in the $xy$ plane and a point of inflection in the $xz$ plane. 
However, in the twisted geometry $\rho_{\rm tw}(\bmr) \simeq \phi_{5d_{xz}}(\bmr)\phi_{6p_y}(\bmr)$, and the charge distribution will have nodes in the $xy$, $yz$, and $xz$ planes. One can qualitatively see that $\rho_{\rm tw}(\bmr)$ will be generally be smaller in magnitude and consequently $K_{HL}$ will be smaller (but still greater than zero) at the twisted geometry. 


\section{Permanent dipole moment}
Since the totally symmetric representation ($A_1$) in the $C_{2v}$ point group transforms as the $z$ vector, the static dipole (both in ground and excited states) must lie along the $z$ axis, assigning the direction of the dipole moment found by F\"oller and Marian.\cite{fol17a} To determine its size and sign, the ground state has an electron-rich carbazole and an electron-poor carbene, such that the ground-state dipole points from the $\delta^{-}$ Cz to the $\delta^+$ carbene, shown in \figr{molstruc}(b). In the $S_1$ state the LUMO localized on Au/C acquires electron density whereas the HOMO on Cz loses it, consistent with the ``remarkable'' change in dipole moment computed by F\"oller and Marian.\cite{fol17a} 

\section{Conclusions}
We have seen how using theoretical arguments we can determine the approximate form of the HOMO and LUMO, and explain why the planar geometry is emissive but the twisted geometry not, and why SOC does not easily mix $S_1$ and $T_1$, but can easily mix $S_2$ and $T_1$, confirming previous experimental and theoretical investigations\cite{di17a,fol17a}. We have also determined the direction of the static dipole moments of CMAs and the polarization of the $S_0\to S_1$ and $S_0\to S_2$ transitions, and the irreps of the electronic wavefunctions of $S_1,S_2$ and $T_1$ in the planar and twisted conformations. We have found that the central metal atom plays a crucial role in the chromophore and emission is qualitatively driven by a $6p\to5d$ transition. We have also examined the $S_1/T_1$ energetic separation and shown that this decreases upon twisting the molecule, provided that $S_1$ and $T_1$ are well-described by HOMO$\to$LUMO excitations. 

This analysis can also inform the future design of CMAs. To increase the dipole moment the HOMO and LUMO need more amplitude on the central metal atom---though this comes at the cost of increasing $K_{HL}$. To blue-shift emission the amide should stabilize the $5d_{yz}$ less, which could be achieved by lowering the HOMO of the amide or decreasing the orbital amplitude on the N atom (such that there is still amplitude of the HOMO on Au). Similarly, blue-shifting emission could be achieved by stabilizing the LUMO less with a less electron-donating group than N next to the carbene carbon. 

\section*{Acknowledgements}
TJHH thanks Richard H.\ Friend for valuable discussions and Jesus College, Cambridge for a Research Fellowship.

\bibliography{refbig}
\end{document}